\documentclass[twocolumn,showpacs,preprintnumbers,amsmath,amssymb]{revtex4}
\usepackage{dcolumn}% Align table columns on decimal point
\usepackage{bm}% bold math
\usepackage{hyperref}

\begin{document}

\def\be{\begin{equation}}
\def\ee{\end{equation}}
\def\e{\end{equation}}
\def\({\left(}
\def\){\right)}
\def\[{\left[}
\def\]{\right]}
\def\vp{\vect p}
\def\vx{\vect x}
\def\vy{\vect y}
\def\vz{\vect z}
\def\vect#1{\skew{-1}{\mathaccent"017E}{#1}}

\title{Comment on ``Casimir energies with finite-width mirrors''}

\author{Ignat Fialkovsky}
 \email{ignat.fialk@paloma.spbu.ru}
 \affiliation{Instituto de Fisica, USP,
    R. do Matao, Travessa R, 187. CEP 05508-090, Sao Paulo, Brasil
 \\
 Department of Theoretical Physics, State University of Saint
Petersburg, Saint Petersburg 198504, Russia.}\author{Yuriy Pis'mak}%
 \email{pismak@JP7821.spb.edu}
 \affiliation{Department of Theoretical Physics, State University of Saint
Petersburg, Saint Petersburg 198504, Russia.}%Lines break automatically or can be forced with \\
\author{Vladimir Markov}
 \email{markov@thd.pnpi.spb.ru}
\affiliation{Department of Theoretical Physics, Petersburg Nuclear
Physics Institute, Gatchina 188300, Russia.
}%

\begin{abstract}
We comment on a recent publication \cite{Fosco Lombardo Mazzitelli 08}
on Casimir energies
for material slabs (`finite width mirrors')
and report a discrepancy between results obtained there for a single mirror
and some previous calculations. We provide a simple consistency check
which proves that the method used in [1] is not reliable when applied to
approximations of piecewise constant profile of the mirror.

We also present an alternative method for calculation of the
Casimir energy in such systems based on \cite{Pismak Fialkovsky
08}. Our results coincide both with perturbation theory and with
some older \cite{Bordag'95} and more recent \cite{Vass}
calculations, but differ from those of \cite{Fosco Lombardo Mazzitelli 08}.

\end{abstract}


\maketitle

In a recent publication on Casimir energies and interaction of
material slabs (`finite width mirrors') \cite{Fosco Lombardo
Mazzitelli 08} there was presented a thorough and rather general
treatment of the problem at hand. To model the presence of matter
in the system of massless scalar field the authors of \cite{Fosco
Lombardo Mazzitelli 08} introduce into the action an additional
term concentrated in a given domain of the space-time. In the
framework of renormalizable quantum field theory  this method was
initially proposed by Symanzik \cite{Symanzik'81} though his paper
is not cited in \cite{Fosco Lombardo Mazzitelli 08}. Such models
with position dependent mass terms (also called defects) were
studied extensively in recent years, in particular in the case of
delta-potentials which effectively describe thin films, see for
instance \cite{FMP 03-08}, \cite{Milton 0401}. Surprisingly the
problem of interaction of finite width mirrors hasn't  been
investigated in full, despite several attempts e.g.
\cite{Bordag'95}, \cite{Bordag'98}, \cite{Feinberg'01}.

The Casimir energy for massive scalar field interacting with a
homogenous plane slab of finite width has been calculated quite a
long time ago by Bordag \cite{Bordag'95}, and also recently
rederived in \cite{Vass} by Vassilevich and
Konoplya
\footnote{
 By their titles these papers are seemingly not connected with the problem at hand
but still both of them contain explicit calculation of the Casimir energy of a plane slab.}.
A similar problem is considered in Section {III.B} of
\cite{Fosco Lombardo Mazzitelli 08}. A smooth approximation of
piecewise constant potential is investigated there and Casimir
energy for the case when this limit is approached is obtained.
However the outcome presented there  (equation (68)) disagree  with
(well defined) massless limit of the above mentioned results,  and
in \cite{Fosco Lombardo Mazzitelli 08} there is no any discussion
of this discrepancy.

In \cite{FMP 03-08} we elaborated a calculation approach for
singular potentials describing the interaction of
thin material films with fields of quantum electrodynamics.
In order to generalize it to the case of
3-dimensional defects we studied recently the model of massive
scalar field interacting with a single plane slab \cite{Pismak
Fialkovsky 08}.
We developed a simple calculation method that differs both from
\cite{Bordag'95, Vass} and also from one employed in  \cite{Fosco
Lombardo Mazzitelli 08}. We calculated the propagator and the
Casimir energy \cite{Pismak Fialkovsky 08}. In appropriate
regularization the later result coincides with ones calculated in
\cite{Bordag'95, Vass} (but differs from that of \cite{Fosco
Lombardo Mazzitelli 08}). It is also in agrement with usual
perturbation theory both for the massive case and in the massless
limit.

One could argue that the discrepancy between results of
\cite{Pismak Fialkovsky 08, Bordag'95, Vass} and \cite{Fosco
Lombardo Mazzitelli 08} originates from the smooth approximation
used in \cite{Fosco Lombardo Mazzitelli 08} to describe singular
(non-continuous) potential. Similar problems have been discussed
in the framework of Dirac equation perturbed by a delta-function
potential (e.g. see \cite{McKellar'87}). However, in our case
it is not the reason of the above mentioned discrepancy. As we show
below, the calculation method proposed in \cite{Fosco Lombardo
Mazzitelli 08} is not reliable when applied to the case of
(approximated) piecewise constant potential, and the final
expression (68)\cite{Fosco Lombardo Mazzitelli 08} is to be
reconsidered.

In Section I of this Comment we present a consistency check of the calculation
method developed in \cite{Fosco Lombardo Mazzitelli 08},
while in Section II we sketch our own approach to similar problem.

\section{Consistency check}

The calculation approach of \cite{Fosco Lombardo Mazzitelli 08} is
based on expressing the Casimir energy as a trace of a logarithm
of an integral operator, and this step is widely used in Casimir
calculations. The trace of the operator can be calculated directly
as a sum of its eigenvalues. For their determination authors of
\cite{Fosco Lombardo Mazzitelli 08} propose particular
Sturm-Liouville problem (on a finite interval) which is obtained
using a specific coordinate transformation, see Section II.A
\cite{Fosco Lombardo Mazzitelli 08}.

As a consistency check of this method we use  it to calculate the
trace of the integral operator $D(z,z')$, defined in (32) of
\cite{Fosco Lombardo Mazzitelli 08}, and compare it with result of a
straightforward alternative calculation. Expressing the trace of
$D(z,z')$ as a sum of its eigenvalues $\mu_l$ and
using (33) \cite{Fosco Lombardo Mazzitelli 08} we obtain
\be
 \textrm{Tr} D(z,z')
    = \sum_l \mu_l=\sum_l \frac{\alpha_l-1}
        {\tilde\lambda(\omega,\textbf{k}_\parallel)}.
    \label{Tr_mu}
\ee
For the limit of piecewise constant profile being approached,
$\alpha_l$ is given in (66) \cite{Fosco Lombardo Mazzitelli 08}
\be
    \alpha_l=\alpha_l(\omega,k_\parallel, \epsilon)
         = 1+\frac{2\epsilon\tilde{\lambda}(\omega,\textbf{k}_\parallel)}
            {l^2\pi^2+(2\epsilon\kappa)^2}.
\label{alph}
\ee
Then one easily finds
\be
 \textrm{Tr} D(z,z')
    = \sum_{l=1}^\infty \frac{2\epsilon}{l^2\pi^2+(2\epsilon\kappa)^2}
    =\frac{2\epsilon\kappa\coth(2\epsilon\kappa)-1}{4\epsilon \kappa^2}
    \label{Tr a}.
\ee

On the other hand, it is straightforward to calculate the trace
of $D$ without appealing to any eigenvalue problem
\begin{eqnarray}
 \textrm{Tr} D(z,z')
    &=& \int_{-1/2}^{1/2} dz D(z,z)=\nonumber\\
    &=&\int_{-\infty}^{\infty} dx D(x,x)\sigma_\epsilon(x)
        =\frac{1}{2\kappa}.
    \label{Tr b}
\end{eqnarray}
Evidently, (\ref{Tr b}) contradicts  to (\ref{Tr a}).
At the same time it is straightforward to prove the equivalence of
the trace definitions used in (\ref{Tr a}) and (\ref{Tr b}) taking
into account that  $\psi_\alpha$  (33) \cite{Fosco Lombardo
Mazzitelli 08} constitute a complete set of eigenfunctions of
Sturm-Liouville problem (39) \cite{Fosco Lombardo Mazzitelli 08}.
Moreover, it must be emphasized here that (\ref{Tr b}) is valid
independently of any particular profile of $\sigma_\epsilon(x)$
provided it satisfies the normalization condition (7)[1]. It is
due to the fact that $D(z,z)$ (or equivalently $D(x,x)$) is
position-independent.

We see that the spectral problem used for calculation of the
eigenvalues of $D$ in the case of (approximated) piecewise
constant profile is either ill posed, or not equivalent to the
original problem. It gives a wrong answer (\ref{Tr a}) for the
trace of $D$. In virtue of (31) \cite{Fosco Lombardo Mazzitelli
08} the same must also hold for the trace of the operator $\ln
\tilde{{\cal K}}$ which defines the Casimir energy (30)
\cite{Fosco Lombardo Mazzitelli 08}. Hence, it is not legitimate
to use the eigenvalues $a_l$ (\ref{alph}) for the (correct) calculation of
the Casimir energy (43) \cite{Fosco Lombardo Mazzitelli 08}.

Thus, we must conclude that the calculation method proposed in
\cite{Fosco Lombardo Mazzitelli 08} contains an internal
inconsistency when applied to (approximations of) piecewise
constant profile of the mirror. One can also check that an
expansion of (67) \cite{Fosco Lombardo Mazzitelli 08} in a power
series in constant $\lambda$ (acknowledged in Section III.C
\cite{Fosco Lombardo Mazzitelli 08}) does not coincide with the
usual perturbation theory.

These arguments unambiguously show that
the result for the Casimir
energy of a single slab (68) \cite{Fosco Lombardo Mazzitelli 08}
is presumably incorrect.

\section{Alternative approach}
In \cite{Pismak Fialkovsky 08} we presented a detailed treatment
of the similar system of piecewise constant profile without
reference to more general cases. We restricted ourselves to
consideration of homogenous and isotropic infinite plane slab of
thickness $\epsilon$, placed in the $x_1x_2$ plane.

Casimir interaction of (multi) layered systems has been actively
studied for dielectric materials (e.g., within surface modes
formalism \cite{modes}, or macroscopical field operators method
\cite{fields}). However, the mathematical formulation of these
problems differ from one considered here, and no direct comparison
of the results is possible.

For modeling of the interaction of massive scalar field with
volume defects we followed the Symanzik approach
\cite{Symanzik'81}. The complete action of the model is following
\begin{eqnarray*}
   &S=&S_0+S_I+J \phi\\
  &&S_0=\frac12\int d^4x\, (\partial_\mu\phi(x)\partial_\mu\phi(x)+m^2 \phi^2(x)),\\
  &&S_I= \frac{\lambda}2\int d^4x\, \theta(\epsilon, x_3)\phi^2(x).
\end{eqnarray*}
The distribution function $\theta(\epsilon, x_3)$ represents the
piecewise constant profile being equal to $1/\epsilon$ when
$|x_3|<\epsilon/2$, and zero otherwise. Then there is no need for
introduction of implicit variables' change defined by (21)
\cite{Fosco Lombardo Mazzitelli 08}, and we can proceed explicitly
in Cartesian coordinates.

We consider the generating functional of Green's functions and
similarly to \cite{Bordag'98} introduce auxiliary fields defined
on the support of the defect (this line is also followed in
\cite{Fosco Lombardo Mazzitelli 08} with minor generalizations).
These fields satisfy free boundary conditions on the edges of the
layer (in other words there are no constrains imposed
on the fields). Following this approach one introduces integral operators
acting on the auxiliary fields with support in the finite
interval, such as $\lambda^{-1}$ in (12) of \cite{Fosco Lombardo
Mazzitelli 08}. In our case it reduces to a unity operator and
there is no need for any special treatment of its boundary
conditions.

Functional integral takes then explicitly gaussian form and we are
able to derive both the Casimir energy and the propagator,
arriving at
\be
\begin{array}{c}
  \mathcal{Z}[J]=({\rm Det} Q)^{-1/2}\,
    \exp{\left\{\frac12J\hat S J\right\}}, \\
  \hat S=D-\eta (\Delta{\cal O}) Q^{-1}({\cal O} \Delta),\ \
    Q=\textbf{1}+\eta({\cal O} \Delta {\cal O}).   \\
\end{array}
\label{Z(j)}
\ee
here $\eta={\lambda}{\epsilon}^{-1}$. The definitions of $\hat S$ and $Q$
must be understood in terms of integral operators.
With  $\Delta=(-\partial^2+m^2)^{-1}$ we denote the standard free
propagator of scalar field, and the projecting operator ${\cal O}$ acts as
$$
    \psi{\cal O}\phi\equiv
        \int d\vx\int_{-\epsilon/2}^{\epsilon/2} dx_3\psi\phi,
$$
The unity operator $\textbf{1}$ is also defined on the defect only.
We shall note here that
our operator $Q$ is the direct analog of the ${\cal K}$
(24) \cite{Fosco Lombardo Mazzitelli 08}. However, the trace calculation which
defines the Casimir energy and is presented below, differs significantly
in our approach.

The Casimir energy density per unit area of the layer is given
by
\be
{\cal E}
    =\int\frac{d^{3}\vp}{2(2\pi)^{3}}
        \mbox{Tr}\ln[Q(\vp; x_3,y_3)],
    \label{E_TrLn2}
\e
where the Fourier transformation of the  coordinates
parallel to the defect (i.e.  $\vx=(x_0, x_1, x_2)$) was performed,
$\vp=(p_0,p_1,p_2)$.

Introducing operator $U\equiv Q^{-1}-1$ and using the definition
of $Q$ we can express the $\eta$-derivative
of the integrand of (\ref{E_TrLn2}) in the following form
$$
    \partial_\eta \ln Q=-\frac{U}{\eta}.
$$

For explicit calculation of $U$ we note
that it is proportional to a Green function
of an ordinary differential operator:
\begin{eqnarray}
  &&K_\rho U= -\eta   \label{s1} \\
  && K_{V}(x,y)\equiv
        \left(-\frac{\partial^2}{\partial x^2}+V^2\right)\delta(x-y).\nonumber
\end{eqnarray}
where $\rho=\sqrt{\eta+E^2}$, $E=\sqrt{\vp^2+m^2}$.

Employing the symmetry conditions $U(x,y)=U(y,x)=U(-x,-y)$ which follow
from the definition of $U$ we can write
\be
U=-\eta \frac{e^{- \rho |x-y|}}{2\rho}
    +2a\cosh\((x+y) \rho\)
    +2b\cosh\((x-y) \rho\).
\ee
From (\ref{Z(j)}) and definition of $U$ it follows
that
$$
U+\eta {\cal O}\Delta {\cal O}(1+U)=0.
$$
Then one can derive the coefficients $a$, $b$ as
\begin{eqnarray}
  a&=&-\frac{\xi \eta^2 e^{\epsilon {\rho}}} {2 {\rho}},\qquad
    b=-\frac{\xi \eta (E-{\rho})^2}{2  \rho},
    \label{abc}\\
  &&\xi =\(e^{2\epsilon \rho}(E+ \rho)^2-(E- \rho)^2\)^{-1}.\nonumber
\end{eqnarray}

For the energy density we have
\be
\label{res1}
{\cal E}
    =-\mu^{3-d}
        \int\frac{d^{d}\vp}{2(2\pi)^{d}}\int_0^\eta \frac{d\eta}{\eta}\,
        \mbox{Tr} U.
\ee
where we introduced dimensional regularization to free oneself
from ultraviolet divergencies ($d=3$ corresponds to removing of
regularization), and  an auxiliary mass parameter $\mu$. We have
chosen the lower limit of integration over $\eta$ to satisfy the
energy normalization condition ${\cal E}|_{\eta=0}=0$. It can be
shown that the integral is convergent at $\eta=0$.

The calculation of trace of the operator $U$ is straightforward
\be
\mbox{Tr} U\equiv \int_{-\epsilon/2}^{\epsilon/2} dx U(x,x)
    =2 b\epsilon+\frac{4 a \sinh(\epsilon \rho)-\epsilon\eta}{2
    \rho}.
    \label{Tr U}
\ee

Next, putting (\ref{abc}) into (\ref{Tr U}) we can prove directly that
\be
\mbox{Tr} U
    =-\eta\frac{\partial}{\partial\eta}
        \ln\left[ \frac{e^{-\epsilon(E+\rho)}}{4E\rho\xi}\right].
    \label{Tr U1}
\ee
Thus, from (\ref{res1}) and  (\ref{Tr U1}) we obtain
the following expression for the Casimir energy
\begin{widetext}
\be
{\cal E}=\mu^{3-d}\int \frac{d^{d}\vp}{2 (2\pi)^{d-1}}
    \ln\[
        \frac{e^{- \epsilon E}}{4 E \rho}
            \(e^{ \epsilon \rho }(E+ \rho)^2-e^{- \epsilon \rho }(E- \rho)^2\)
    \],\quad \rho=\sqrt{E^2+\lambda\epsilon^{-1}}
    \label{E}
\ee
\end{widetext}
It can be shown that (\ref{E})  is in full agreement
with the usual perturbation theory.

It is easily to generalize (\ref{E}) for non-local translation
invariant $\lambda\equiv\lambda(\vx-\vy)$ as considered in
\cite{Fosco Lombardo Mazzitelli 08} (see equation (4) there). In this
case the final expression for ${\cal E}$ (\ref{E}) remains the
same provided that $\lambda$ is replaced there with $\tilde\lambda
(\vp)$ -- Fourier image of $\lambda(\vx)$ defined in (26)
\cite{Fosco Lombardo Mazzitelli 08}. In the massless model (with
$m=0$, $E=|\vp|$) the energy ${\cal E}$ obtained in such a way can
be compared with $\mu^{3-d}{\cal E}_0$ for ${\cal E}_0$ presented
in (67) \cite{Fosco Lombardo Mazzitelli 08} for
plane slab. One can easily check that these results
do not coincide.

To compare (\ref{E}) with results of \cite{Bordag'95} and
\cite{Vass} one needs to replace the zeta-function regularization
used there with dimensional one applied in our approach. Using
(15) of \cite{Bordag'95} one can obtain a generalization of (8)
\cite{Bordag'95} (written for $d=3$) to $d$-dimensional case
in the form
\be
V_{\textit{eff}}
    =
        \frac{\Omega_d \mu^{3-d}}{2(2\pi)^{d} }
        \int_m^\infty dk (k^2-m^2)^{d/2}
            \frac{\partial}{\partial k}\ln s_{11}(ik)
        \label{Veff}
\ee
where $\Omega_d=2\pi^{d/2}/\Gamma(d/2)$ is the volume of the
$(d-1)$-dimensional sphere in the $d$-dimensional space, and
$s_{11}(ik)$ for considered case of plane slab is defined by (22)
\cite{Bordag'95}. Integrating by part in (\ref{Veff}) and changing
integration variable $k=\sqrt{|\vp|^2+m^2}$, one can check that the
right hand sides of (\ref{E}) and (\ref{Veff}) coincide (up to
change of notation $L=\epsilon$, $V_0=\eta$). In a similar way
one can verify that result obtained in \cite{Vass} for the plane
slab for $d=2$ with help of zeta-function regularization agrees
with (\ref{E}).

\vskip 3mm
We hope that the arguments given above are sufficient to conclude
that the calculation methods proposed in \cite{Fosco Lombardo
Mazzitelli 08} need to be thoroughly verified in order to expose
and eliminate theirs defects. Then it may be possible to employ effectively
the basic ideas of \cite{Fosco Lombardo Mazzitelli 08}
in calculation of Casimir energy in the models of
quantum field theory with nontrivial background.

\begin{acknowledgments}
V.M. and Yu.P. are grateful to the Russian Foundation of Basic Research  for
financial support in framework of the grant RFRB $07$--$01$--$00692$. The work
of I.F. was partly supported by FAPESP.

\end{acknowledgments}


\begin{thebibliography}{99}
\bibitem{Bordag'95}
    M.~Bordag,
    \textit{Vacuum energy in smooth background fields},
    J.\ Phys.\ A  {\bf 28}, 755 (1995). Note that the author uses the term
    \textit{effective potential }for the quantity which we identify as Casimir energy.
\end{thebibliography}
\end{document}